\documentclass[aps,prb,twocolumn,groupedaddress,showpacs]{revtex4}

\usepackage[dvipdfm]{graphicx}
\usepackage{bm}

\begin{document}

\title{Berry phase of non-ideal Dirac fermions in topological insulators}

\author{A. A. Taskin}
\author{Yoichi Ando}

\affiliation{Institute of Scientific and Industrial Research,
Osaka University, Ibaraki, Osaka 567-0047, Japan}

\begin{abstract}

A distinguishing feature of Dirac fermions is the Berry phase of $\pi$
associated with their cyclotron motions. Since this Berry phase can be
experimentally assessed by analyzing the Landau-level fan diagram of the
Shubnikov-de Haas (SdH) oscillations, such an analysis is widely
employed in recent transport studies of topological insulators to
elucidate the Dirac nature of the surface states. However, the reported
results have usually been unconvincing. Here we show a general scheme
for describing the phase factor of the SdH oscillations in realistic
surface states of topological insulators, and demonstrate how one could
elucidate the Dirac nature in the real experimental data.

\end{abstract}

\pacs{73.25.+i, 73.20.At, 71.70.Di, 72.20.My}


\maketitle

\section{Introduction}

During the last three decades, the Berry phase \cite{Berry} has become
an important concept in condensed matter physics, \cite{S-W} playing a
fundamental role in various phenomena such as electric polarization,
orbital magnetism, anomalous Hall effects, {\it etc.} \cite{Niu2010} The
Berry phase (or geometrical phase) in solids is determined by
topological characteristics of the energy bands in the Brillouin zone
(BZ) and represents a fundamental property of the system. For example, a
non-zero Berry phase, which can be measured directly in the
magnetotransport experiments, reflects the existence of a singularity in
the energy bands such as a band-contact line in three-dimensional (3D)
bulk states or a Dirac point in a two-dimensional (2D) surface state.
\cite{Mikitik1999} Also, the Berry phase of $\pi$ is responsible for the
peculiar ``anti-localization" effects in carbon nanotubes or graphene.
\cite{TAndo} Recently, the $\pi$ Berry phase has been observed in the
Shubnikov-de Haas (SdH) oscillations in graphene, \cite{N-G2005,Kim2005}
giving one of the key evidences for the Dirac nature of quasiparticles
in the 2D carbon sheet.

The 3D topological insulator (TI) also supports spin polarized 2D Dirac
fermions on its surface,\cite{HK} which can be distinguished from
ordinary charge carriers by a non-zero Berry phase. Recently, several
groups have reported observations of the SdH oscillations coming from
the 2D surface states of TIs.
\cite{BiSb_amro,Ong2010,Fisher_np,BTS_Rapid,Xiong,Morpurgo,HgTe,Bi2Te3nano}
In those studies, a finite Berry phase has been reported, but it usually
deviates from the exact $\pi$ value. For example, in the new TI material
Bi$_{2}$Te$_{2}$Se (BTS), \cite{BTS_Rapid} where a large contribution of
the surface transport to the total conductivity has been observed, the
apparent Berry phase extracted from the SdH-oscillation data was
0.44$\pi$. So far, the Zeeman coupling of the spin to the magnetic field
has been considered \cite{Fisher_np} as a possible source of such a
discrepancy. Here, we show that in addition to the
Zeeman term, the deviation of the dispersion relation $E(k)$ from an
ideal linear dispersion \cite{DasSarma2010} can shift the Berry phase
from $\pi$. We further show how the real experimental data for non-ideal
Dirac fermions could be understood by taking into account those
additional factors.

\section{energy dispersion of surface states}

\begin{figure}\includegraphics*[width=8.0cm]{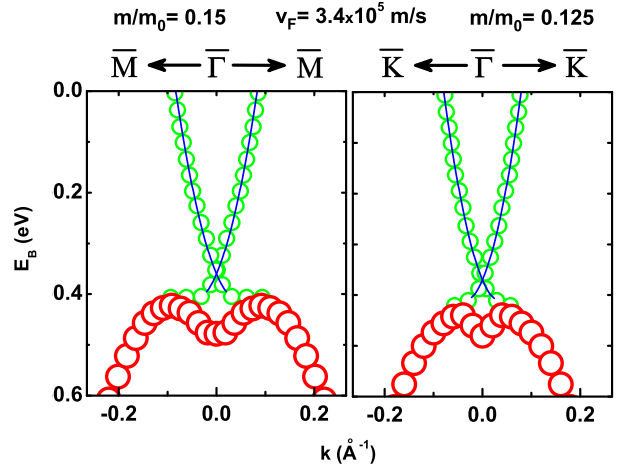}
\caption{(Color online) 
Experimental band dispersions (symbols) in Bi$_{2}$Te$_{2}$Se measured by ARPES in Ref. 
\onlinecite{BTS_Hasan} and the fitting of Eq. (1) to the surface state (solid line).
Large symbols depict the bulk state.
} 
\label{fig1}
\end{figure}

The energy dispersion of the surface states in TIs can be directly measured in 
angle-resolved photoemission spectroscopy (ARPES) experiments. 
As an example, Fig. 1 shows the dispersion of the surface state (together with
the bulk state) in BTS reported by 
Xu {\it et al.} \cite{BTS_Hasan}
One can easily recognize that $E(k)$ is not an ideal 
Dirac-like dispersion, but it can be fitted reasonably well for the 
two high-symmetry axes with 
\begin{equation}
E(k) = v_{F} \hbar \, k + \frac{\hbar^{2}}{2m} k^{2},
\end{equation}
with a single Fermi velocity $v_{F}$ = 3.4$\times$10$^{5}$ m/s and the
effective mass $m$ which slightly varies with the direction in the
surface BZ as shown by the solid lines in Fig. 1 [$m/m_{0}$ = 0.15
(0.125) for the $\bar{\Gamma} \rightarrow \bar{M}$ ($\bar{\Gamma}
\rightarrow \bar{K}$) direction with $m_0$ the free electron mass].

Similar fittings can be obtained for other TIs owing to the progress in
the ARPES studies of these materials. \cite{HK,HgTe,H5,Chen}

\section{Berry phase in quantum oscillations}

It is commonly accepted that quantum oscillations observed in 3D metals
can be well understood within Lifshits-Kosevich \cite{L-K} (the de
Haas-van Alphen effect) and Adams-Holstein \cite{A-H} (the SdH effect)
theories. Recently this approach has been generalized to describe
magnetic oscillations in graphene, which is a 2D system with a
Dirac-like spectrum of charge carriers. \cite{Sh-G-B, G-Sh} There are
two most prominent features that distinguish such systems from materials
with a parabolic spectrum: First, rather weak magnetic fields are
sufficient to bring the system into a regime where only a few Landau
levels are occupied. Second, Dirac quasiparticles acquire the Berry
phase of $\pi$ in the cyclotron motion, changing the phase of quantum
oscillations.

In the SdH effect, the oscillating part of $\rho_{xx}$ follows
\begin{equation}
\Delta \rho_{xx} \sim \cos[2\pi (\frac{F}{B}-\gamma)],
\end{equation}
where $F$ is the oscillation frequency and 2$\pi \gamma$ is the phase factor
($0 \le \gamma < 1$). 
This is the same $\gamma$ as in the Onsager's semiclassical quantization 
condition \cite{Shoenberg1984}
\begin{equation}
A_{N}=\frac{2\pi e}{\hbar}B(N+\gamma),
\end{equation}
when the $N$-th Landau level (LL) is crossing the Fermi energy $E_F$
($A_{N}$ is the area of an electron orbit in the $k$-space).
$\gamma$ is directly related to the Berry phase through\cite{Mikitik1999}
\begin{equation}
\gamma - \frac{1}{2} = -\frac{1}{2\pi} \oint_{\Gamma} \vec{\Omega}\,d\vec{k}, 
\end{equation}
where $\vec{\Omega}(\vec{k})$=i$\int d\vec{k} \, u_{\vec{k}}^{*}(\vec{r}) \, 
\vec{\nabla}_{\vec{k}} u_{\vec{k}}(\vec{r})$ is the Berry connection, 
$u_{\vec{k}}(\vec{r})$ is the amplitude of the Bloch 
wave function, $\Gamma$ is a closed electron orbit (the intersection of the Fermi 
surface $E(\vec{k}) = E_F$ with the plane $k_{z} = const$). For spinless quasiparticles,
it is known \cite{Mikitik1999,Shoenberg1984} that the Berry phase is zero for 
a parabolic energy dispersion ($\gamma$ = $\frac{1}{2}$) and $\pi$ for a linear 
energy dispersion ($\gamma$ = 0).

Experimentally, $\gamma$ can be obtained from an analysis of the
Landau-level (LL) fan diagram. There are three quantities which are
often used as abscissa for plotting a LL fan diagram: (i) Landau level
index $N$, which determines the energy $E_{N}$ of the $N$-th LL. (ii)
Filling factor $\nu$ ($ \equiv \frac{N_{s}S}{N_{\phi}}$, where $N_{s}$
is the density of charge carriers, $S$ is the area of the sample,
$N_{\phi}$ = $\frac{BS}{\Phi_{0}}$ is the number of flux quanta, and
$\Phi_{0}$=$\frac{h}{e}$ is the flux quantum). (iii) An integer number
$n$ which marks the $n$-th minimum of the oscillations in $\rho_{xx}$.
Although all three quantities are related to each other, the most
straightforward way to plot a LL fan diagram from the $\rho_{xx}$
oscillations in a 2D system \cite{Ong2010} is to assign an integer $n$
to a minimum of $\rho_{xx}$ (or a half-integer to a maximum of
$\rho_{xx}$). From Eq. (2), one can see that the first minimum in
$\rho_{xx}$ is always in the range of $0 < \frac{F}{B_{1}} \leq 1$.
Thus, the plot of $F/B_{n}$ vs $n$, which makes a straight line with a
unit slope for periodic oscillations, is uniquely defined and cuts the
$n$-axis between 0 and 1 depending on the phase of the oscillations,
$\gamma$.

The ordinate 1/$B_{n}$ in a LL fan diagram is determined by the Landau
quantization of the cyclotron motion of electrons in a magnetic field.
In 2D systems, upon sweeping $B$, $\rho_{xx}$ shows a maximum (or a
sharp peak in the quantum Hall effect \cite{Ong2010}) each time when
$E_{N}(B)$ crosses the Fermi level. Thus, the position of the maximum in
$\rho_{xx}$ that corresponds to the $N$-th LL, 1/$B_{N}$, is given by
\begin{equation}
2\pi \Big(\frac{F}{B_{N}}-\gamma \Big) = 2\pi N.
\end{equation}
On the other hand, the $n$-th minimum in $\rho_{xx}$ occurs at 1/$B_n$ when 
$2\pi(\frac{F}{B_n}-\gamma) = 2\pi n - \pi$, so the positions of the maxima 
and minima are shifted by $\frac{1}{2}$ on the $n$-axis.

The Onsager's relation \cite{Shoenberg1984} gives $F$ in terms of the Fermi wave 
vector $k_{F}$ as
$F = (\hbar  / 2\pi e)\pi k_{F}^{2}$, and this $k_{F}$ can be calculated from Eq. (1) as
\begin{equation}
k_{F}^{2} = 2\Big(\frac{m v_{F}}{\hbar}\Big)^{2}\Bigg(1+\frac{E_{F}}{m v_{F}^{2}} - 
\sqrt{1+\frac{2E_{F}}{m v_{F}^{2}}} \Bigg).
\end{equation}
Also, when $E_F$ is at the $N$-th LL, there is a relation
\begin{equation}
E_{N}(B_{N}) = E_{F}.
\end{equation}
From Eqs. (5)--(7), one obtains
\begin{equation}
\gamma = \frac{m v_{F}^{2}}{\hbar \, \omega_{c}}\Bigg(1+\frac{E_{N}}{m v_{F}^{2}} - 
\sqrt{1+\frac{2E_{N}}{m v_{F}^{2}}} \Bigg) - N,
\end{equation}
where $\omega_{c}$=$eB/m$ is the cyclotron frequency. 

In general case, $\gamma$ is a function of $B$, meaning that
oscillations in $\rho_{xx}$ are quasi-periodic in $1/B$. 
In order to calculate $\gamma$ one needs to find the eigenvalues $E_{N}$ 
for a given Hamiltonian.

\section{Model Hamiltonian}

For the (111) surface state of the Bi$_{2}$Se$_{3}$-family TI compounds,
the Hamiltonian for non-ideal Dirac quasiparticles in perpendicular
magnetic fields can be written as \cite{MQOsc}
\begin{equation}
\hat{H} = v_{F}(\Pi_{x}\sigma_{y}-\Pi_{y}\sigma_{x}) + 
\frac{{\bf\Pi}^{2}}{2m} - \frac{1}{2} \,g_{s} \mu_{B}B \sigma_{z} ,
\end{equation}
where the Landau gauge ${\bf A} = (0,By,0)$ for the vector potential is
used, ${\bf \Pi}$=$\hbar \, {\bf k}$+$e {\bf A}$, $\sigma_{i}$ are the Pauli 
matrices, $\mu_{B}$ is the Bohr magneton, and $g_{s}$ is the surface $g$-factor.
The LL energies are given by \cite{MQOsc,PP2009}
\begin{equation}
E_{N}^{(\pm)} = \hbar \omega_{c} N \pm \sqrt{2 \hbar \, v_{F}^{2} e B N + 
\Big(\frac{1}{2} \hbar \omega_{c} - \frac{1}{2} \,g_{s} \mu_{B} B\Big)^{2}} ,
\end{equation}
where ``$+$" and ``$-$" branches are for electrons and holes, respectively. 
The obtained eigenvalues $E_{N}$ define the exact positions of maxima in 
$\rho_{xx}$ and, thus, the phase of oscillations through Eq. (8).

In two extreme cases, for non-magnetic fermions ($g_{s}$ = 0), 
Eq. (8) gives the expected results. First, for a
linear dispersion (ideal Dirac fermions), $m \rightarrow \infty$ leads to 
$E_{N}$=$\pm \sqrt{2\hbar \, e v_{F}^{2} B N}$ and 
$\gamma \rightarrow \frac{E_{N}^2}{2\hbar \, e v_{F}^{2} B} - N$, 
giving $\gamma$ = 0 (Berry phase is $\pi$). 
Second, for a parabolic dispersion, $v_{F} \rightarrow$ 0 leads to 
$E_{N}$=$\hbar \,\omega_{c}(N+\frac{1}{2})$ and
$\gamma \rightarrow \frac{E_{N}}{\hbar \, \omega_{c}} - N$, 
giving $\gamma$ = $\frac{1}{2}$ (Berry phase is zero). 
This gives confidence that the expression for $\gamma$ given in Eq. (8) is
generally valid for the topological surface state with a non-ideal Dirac
cone described by Eq. (1).

\section{Landau-level fan diagram for non-ideal Dirac fermions}

\begin{figure}\includegraphics*[width=6.5cm]{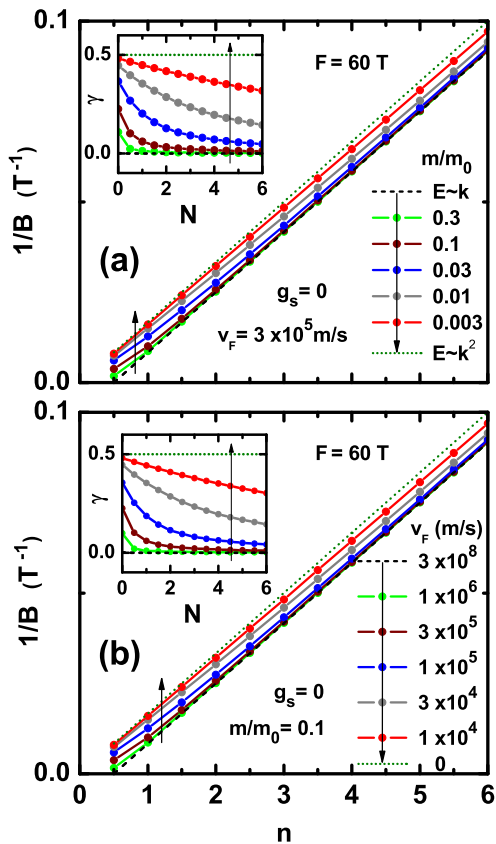}
\caption{(Color online) 
(a) Landau level fan diagram calculated for $F$ = 60 T,
$v_{F}$ = 3$\times$10$^{5}$ m/s, $g_{s}$ = 0, and different $m/m_{0}$.
Arrows show the direction of decreasing $m/m_{0}$. The dashed and dotted
lines are the expected behaviors for an ideal Dirac dispersion and a
parabolic dispersion, respectively. 
(b) Landau level fan diagram calculated for $F$ = 60 T, $m/m_{0}$ = 0.1,
$g_{s}$ = 0, and different $v_{F}$. Arrows show the direction of
decreasing $v_{F}$. Insets show the calculated
$\gamma(N)$.
} 
\label{fig2}
\end{figure}

Let us first consider how the LL fan diagram will be modified, when both
linear and parabolic terms are present in the Hamiltonian [Eq. (9)]. For
the moment, the Zeeman coupling of the electron spin to the magnetic
field is assumed to be negligible ($g_{s}$ = 0). Figure 2 (a) shows the
calculated positions of maxima and minima in $\rho_{xx}$ for
oscillations with $F$ = 60 T and $v_{F}$ = 3$\times$10$^{5}$ m/s as
$m/m_{0}$ is varied. One can see that upon decreasing $m/m_{0}$, the
calculated lines on the LL fan diagram are gradually shifting upward
from the ideal Dirac line that crosses the $n$-axis at exactly
$\frac{1}{2}$. Moreover, the lines are not straight anymore, which is
clearly inferred in the dependence of $\gamma$ vs $N$ shown in the
inset. With decreasing $N$ (increasing $B$), $\gamma$ becomes larger,
reflecting the change in the phase of oscillations at high fields.

Similar change in the LL fan diagram occurs if we modify another parameter, 
$v_{F}$. As shown in Fig. 2 (b), the calculated lines are gradually shifting 
upward from the ideal Dirac line as $v_{F}$ is decreased.
The results shown in Figs. 2 can be understood as a competition 
between linear and quadratic terms in the Hamiltonian [Eq. (9)]. 
Note that for the whole range of the parameters $v_{F}$ and $m/m_{0}$, the positions 
of maxima and minima in $\rho_{xx}$ lie between two straight lines 
(shown as dotted and dashed lines in Figs. 2) 
corresponding to $\gamma$ = 0 and $\gamma$ = $\frac{1}{2}$.

\begin{figure}\includegraphics*[width=6.5cm]{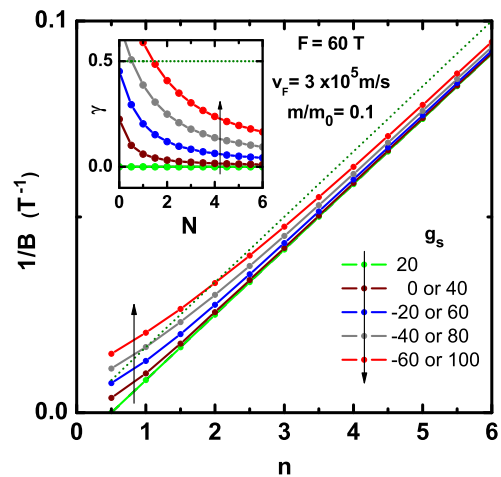}
\caption{(Color online) 
Landau level fan diagram calculated for $F$ = 60 T and different $g_{s}$, 
keeping $v_{F}$ = 3$\times$10$^{5}$ m/s and $m/m_{0}$ = 0.1 constant.  
Arrows show the direction of changing $g_{s}$.
The dotted line is the expected behavior for a parabolic dispersion.
Inset shows the calculated $\gamma(N)$.
} 
\label{fig3}
\end{figure}

Let us now take the Zeeman term into considerations. Figure 3 shows the
LL fan diagram calculated with $F$ = 60 T, $v_{F}$ = 3$\times$10$^{5}$
m/s, and $m/m_{0}$ = 0.1, while $g_{s}$ is varied. To understand the
effect of the Zeeman coupling, it is important to recognize the
following two points: (i) The Zeeman term in Eq. (10) would tend to
cancel the $\frac{1}{2}\hbar\omega_c$ term when $g_s$ is positive. In
fact, when $\frac{1}{2}\hbar\omega_c$ = $\frac{1}{2} g_s \mu_B B$ ({\it
i.e.}, $g_s = 2 m_0/m$) is satisfied, the effect of the finite effective
mass is canceled and the LL fan diagram becomes identical to that for
the linear dispersion (ideal Dirac) case. In the present simulations, we
use $m/m_0$ = 0.1, so that this cancellations occurs when $g_s$ = 20.
(ii) A pair of $g_s$ values that give the same
$|\frac{1}{2}\hbar\omega_c - \frac{1}{2} g_s \mu_B B|$ are effectively
the same in determining the behavior of the LL fan diagram. The result
of our calculations shown in Fig. 3 is a demonstration of these two
points. Since the Zeeman effect is more pronounced at higher fields, the
LL fan diagram in Fig. 3 is strongly modified from a straight line when
the quantum limit is approached, {\it i.e.}, close to $N$ = 0.

\section{The case of BTS}

\begin{figure}\includegraphics*[width=8.7cm]{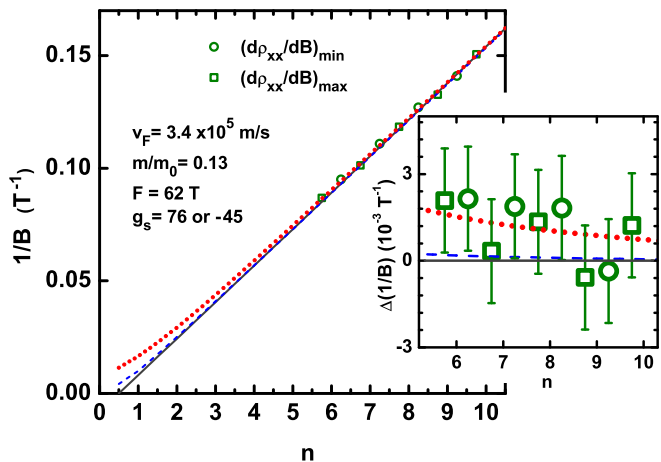}
\caption{(Color online) 
Landau level fan diagram for oscillations in $d\rho_{xx}/dB$ measured at
$T$ = 1.6 K and $\theta \simeq$ 0$^{\circ}$ reported in Ref.
\onlinecite{BTS_Rapid} for BTS. Minima and maxima in $d\rho_{xx}/dB$
correspond to $n+\frac{1}{4}$ and $n+\frac{3}{4}$, respectively. Solid
(dark gray) line is the calculated diagram for an ideal Dirac cone with
$v_{F}$ = 3.4$\times$10$^{5}$ m/s and $F$ = 62 T; dashed (blue) line
includes the effect of the actual dispersion with $m/m_{0}$ = 0.13;
dotted (red) line further includes the Zeeman effect, where $g_{s}$ = 76
or $-45$ was determined from a least-square fitting to the data. Inset
shows the experimental data and calculations after subtracting the contribution
from an ideal Dirac cone, $(1/B)_{\rm Dirac}$, where 
$\Delta (1/B) \equiv (1/B) - (1/B)_{\rm Dirac}$.
} 
\label{fig4}
\end{figure}

Let us examine the real data measured in the BTS sample,
\cite{BTS_Rapid} in the light of the above considerations. 
Figure 4 shows the
LL fan diagram for oscillations in $d\rho_{xx}/dB$ measured at $T$ = 1.6
K in magnetic fields perpendicular to the (111) plane. \cite{BTS_Rapid}
In Ref. \onlinecite{BTS_Rapid}, the data were simply fitted with a
straight line, and the least-square fitting gave a slope of $F$ = 64 T
with the intersection of the $n$-axis at 0.22$\pm$0.12; this result
implies a finite Berry phase, but it was not exactly equal to $\pi$,
which remained a puzzle. \cite{BTS_Rapid} Now, we analyze this LL fan
diagram by considering the non-ideal Dirac dispersion as well as the
Zeeman effect. The ARPES data \cite{BTS_Hasan} for the surface state of
BTS (Fig. 1) gives $v_{F}$ = 3.4$\times$10$^{5}$ m/s and the averaged
effective mass $m/m_{0}$ = 0.13. We fix the oscillation frequency $F$ at
62 T obtained from the Fourier-transform analysis of the $d\rho_{xx}/dB$
oscillations.\cite{BTS_Rapid} 

In Fig. 4, the calculated diagram for an
ideal Dirac cone is shown by the solid (dark gray) line, whereas 
that for the non-ideal Dirac cone with the effective-mass term is shown
by the dashed (blue) line. One can see that the difference is small,
which indicates that the effective mass of 0.13$m_0$ is not light enough
to significantly alter the LL fan diagram. One may also see that these
two lines undershoot the actual data points at smaller $n$, which is even more 
clearly seen in the inset, where the experimental data and the calculations are 
shown after subtracting the contribution from an ideal Dirac cone. 
By further including the Zeeman effect, we can greatly improve the analysis, 
as shown by the dotted (red) line; here, $g_s$ is taken as the only fitting
parameter and a least-square fitting to the data was performed. The best
value of $g_s$ is 76 or $-45$. 

The inset of Fig. 4 makes it clear that it is the slight deviation of
the experimental points from the ideal Dirac line that causes a simple
straight-line fitting of the LL fan diagram to intersect the $n$-axis
not exactly at 0.5. Since the Berry phase in real situations is not a
fixed value but is dependent on the magnetic field, the simple
straight-line analysis of the LL fan diagram should not be employed for
the determination of the Berry phase. Obviously, the SdH oscillations of
the topological surface states are best understood by the analysis which
considers both the the deviation of the energy spectrum of the
Dirac-like charge carriers from the ideal linear dispersion and their
strong coupling with an external magnetic field.

\section{Other materials}

\begin{table}[b]
\centering 
\begin{tabular}{l c c c c} 
\hline\hline 
Material & $v_{F}$ (m/s) & $m/m_{0}$ & Ref. & remark \\ [0.5ex] 
\hline 
Bi$_{2}$Se$_{3}$  & 3.0 $\times$10$^{5}$ & 0.25 & [\onlinecite{H5}] & averaged  \\ 
Bi$_{2}$Te$_{2}$Se & 3.4 $\times$10$^{5}$ & 0.13 & [\onlinecite{BTS_Rapid}] & averaged\\ 
Bi$_{2}$Te$_{3}$ & 3.7 $\times$10$^{5}$ & 3.8 & [\onlinecite{Chen}] & near Dirac point\\ 
graphene & 1 $\times$10$^{6}$ & $\infty$ & [\onlinecite{N-G2005}] & calculations \\ [1ex] 
\hline 
\end{tabular}
\caption{Parameters of the surface states from ARPES.} 
\end{table}

\begin{figure}\includegraphics*[width=8.7cm]{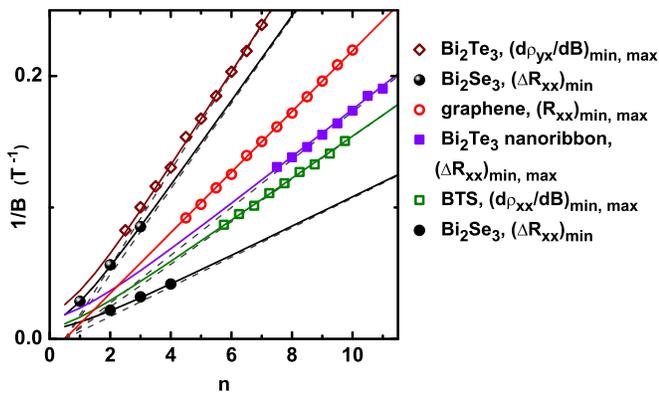}
\caption{(Color online)  
Landau level fan diagrams for SdH oscillations observed in various TIs
and graphene. Symbols are obtained from the published experimental data
in the literature.
Solid lines are calculations taking into account the non-ideal
dispersions of the surface states (determined by $m/m_{0}$) and the
Zeeman coupling to an external magnetic field (determined by $g_{s}$).
Dashed lines are calculations for ideal Dirac fermions ($m/m_{0}$
= $\infty$ and $g_{s}$ = 0). Open diamonds are $(d\rho_{yx}/dB)_{min,max}$
in Bi$_{2}$Te$_{3}$ from Ref. \onlinecite{Ong2010}; filled circles are
$(\Delta R_{xx})_{min}$ in Bi$_{2}$Se$_{3}$ from
Ref. \onlinecite{Fisher_np}; open circles are $(R_{xx})_{min,max}$ in
graphene from Ref. \onlinecite{N-G2005}; filled squares are $(\Delta
R_{xx})_{min, max}$ in a Bi$_{2}$Te$_{3}$ nanoribbon from
Ref. \onlinecite{Bi2Te3nano}; open squares are
$(d\rho_{xx}/dB)_{min,max}$ in BTS from Ref. \onlinecite{BTS_Rapid}.
} 
\label{fig5}
\end{figure}

Similar analysis can be performed for other TIs in which the quantum
oscillations coming from the 2D topological surface states have been
observed. Figure 5 shows the LL fan diagrams for the SdH oscillations
published to date for TI materials,
\cite{Fisher_np,Ong2010,BTS_Rapid,Bi2Te3nano,note1} together with the
data obtained in graphene, \cite{N-G2005} which provides a good
reference for studies of Dirac fermions. We digitized the published
experimental data in the literature and determined ourselves the
positions of minima $1/B_{min}$ and maxima $1/B_{max}$ of the
oscillating parts of resistivity (resistance), Hall resistivity, or 
their derivatives with respect to $B$. The obtained data
for various materials are plotted as functions of $n$ in Fig. 5. Note
that, to avoid ambiguities, we considered only those data that show
oscillations with a single frequency. \cite{note1}

\begin{table}[t]
\centering 
\begin{tabular}{l c c c c} 
\hline\hline 
Material  & Ref. & $F$ (T) & $E_{F}$ (eV) & $g_{s}$ \\ [0.5ex] 
\hline 
Bi$_{2}$Se$_{3}$  & [\onlinecite{Fisher_np}]  & 30.7 & 0.074 & 55 or -39 \\ 
Bi$_{2}$Se$_{3}$  & [\onlinecite{Fisher_np}]  & 88.6 & 0.143 & 55 or -39 \\ 
Bi$_{2}$Te$_{2}$Se & [\onlinecite{BTS_Rapid}]  & 62.0 & 0.152 & 76 or -45 \\ 
Bi$_{2}$Te$_{3}$ & [\onlinecite{Ong2010}]  & 27.3 & 0.074 &65 or -65 \\ 
Bi$_{2}$Te$_{3}$, nanoribbon & [\onlinecite{Bi2Te3nano}]  & 54.7 & 0.101 &65 or -65 \\ 
graphene & [\onlinecite{N-G2005}]  & 43.3 & 0.239 & 0 \\ 
\hline 
\end{tabular}
\caption{Parameters used for the calculations shown in Fig. 5.} 
\end{table}

The parameters of the surface states used in our fan-diagram analyses
have been obtained from the published ARPES data by fitting them in the
same way as for BTS (see Fig. 1). Table I shows $v_{F}$ and $m/m_{0}$
for the Bi$_{2}$Se$_{3}$/Bi$_{2}$Te$_{3}$ family and graphene. These
parameters were fixed during the fitting of the data shown in Fig. 5.
The only parameter that could vary in our calculations was $g_{s}$. Note
that the frequency of oscillations $F$ (and, thus, the Fermi energy
$E_{F}$) is essentially determined by the periodicity of the observed
oscillations. Table II summarizes the parameters thus obtained. The
results of our calculations are shown in Fig. 5 by solid lines. Dashed
lines depict the behavior expected for ideal Dirac cones
($m/m_{0}$=$\infty$) and negligible Zeeman coupling ($g_{s}$ = 0) for
the TI data .
One can clearly see in Fig. 5 that only graphene shows the ideal behavior in 
the LL fan diagram: a straight line that crosses the $n$-axis at 0.5.
All TI materials, despite their essentially Dirac-like nature of the 
surface state, present the LL fan diagrams that deviate from the ideal 
behavior. (The deviations from the dashed lines are most clearly
seen in strong magnetic fields.)

In view of the good agreements between the data and the fittings for
all the materials analyzed in Fig. 5, one may conclude that the advanced
analysis considering both the curvature of the Dirac cone and the Zeeman
effect can reasonably describe the SdH-oscillation data obtained for
TIs and confirm the Dirac nature in their surface states.

\section{Summary}

We derived the formula for the phase $\gamma$ of the SdH oscillations
coming from the surface Dirac fermions of realistic topological
insulators with a non-ideal dispersion given by Eq. (1). We also
calculated how the curvature in the dispersion as well as the effect of
Zeeman coupling affect the Landau-level fan diagram of the SdH
oscillations for realistic parameters. Finally, we demonstrate that the
Landau-level fan diagrams obtained from recently reported SdH
oscillations in topological insulators can actually be understood to
signify the essentially Dirac nature of the surface states, along with a
relatively large Zeeman effect in those narrow-gap materials.

\begin{acknowledgments}
We thank G.P. Mikitik for helpful discussions. 
This work was supported by JSPS (NEXT Program), MEXT
(Innovative Area ``Topological Quantum Phenomena" KAKENHI 22103004),
and AFOSR (AOARD 10-4103).
\end{acknowledgments}


\begin{thebibliography}{99}
\bibitem{note1}
We did not include the data reported in Refs. \onlinecite{BiSb_amro} and
\onlinecite{HgTe}, because the SdH oscillations observed in
Bi$_{0.91}$Sb$_{0.09}$ (Ref. \onlinecite{BiSb_amro}) and in a strained
epitaxial film of HgTe (Ref. \onlinecite{HgTe}) clearly show multiple 
frequencies.
\end{thebibliography}
\end{document}